\documentclass[a4paper, amsfonts, amssymb, amsmath, reprint, showkeys, nofootinbib, twoside]{revtex4-1}

\usepackage[english]{babel}
\usepackage[utf8]{inputenc}
\usepackage[colorinlistoftodos, color=green!40, prependcaption]{todonotes}
\usepackage{amsthm}
\usepackage{mathtools}
\usepackage{physics}
\usepackage{xcolor}
\usepackage{graphicx}
\usepackage[left=23mm,right=13mm,top=35mm,columnsep=15pt]{geometry} 
\usepackage{adjustbox}
\usepackage{placeins}
\usepackage[T1]{fontenc}
\usepackage{lipsum}
\usepackage{csquotes}
\allowdisplaybreaks
\usepackage[pdftex, pdftitle={Article}, pdfauthor={Author}]{hyperref} % For hyperlinks in the PDF
\begin{document}
\title{Cosmological scalar perturbations in Horndeski-like gravity}

\author{Dani de Boe}
    \email{deboe@lorentz.leidenuniv.nl} 
    \affiliation{Institute Lorentz, Leiden University,
PO Box 9506, Leiden 2300 RA, The Netherlands.}

\author{Fabiano F. Santos}
    \email{fabiano.ffs23-at-gmail.com} 
    \affiliation{Instituto de F\'{\i}sica, Universidade Federal do Rio de Janeiro, 21.941-909, Rio de Janeiro, RJ, Brazil.}

\author{Jackson Levi Said}
    \email{jackson.said@um.edu.mt} 
    \affiliation{Institute of Space Sciences and Astronomy, University of Malta, Msida, MSD 2080,
Malta and Department of Physics, University of Malta, Msida, MSD 2080, Malta.}

\begin{abstract}
Scalar-tensor theories are promising dark energy models. A promising scalar-tensor theory, called Horndeski-like gravity, is coming from the application of the Horndeski gravity in string theory and cosmology that takes into account two dilaton fields. In this work we study the stability of the scalar sector of this theory and compare it with that coming from the previously studied tensor sector. With the first-order formalism we investigate the allowed background solutions. Focusing on the background solution with a single scalar field, the entropy coming from particle production $S_{in}$ and that of the apparent horizon $S$ will be studied, which translates into \textit{entropy bounds}. These entropy bounds are compared with the stability of the scalar and tensor sector as well. The gravitational slip (minus one) to entropy ratio is also considered as a possible replacement for the usual shear viscosity to entropy ratio for black holes.
\end{abstract}

\maketitle

\section{Introduction} \label{sec:outline}

In recent years, studies highlight the ongoing efforts to understand scalar perturbations in different gravitational contexts \cite{Becar:2024agj}, from regular and spinning black holes to those influenced by exotic matter like dark fluids \cite{Wagle:2023fwl,Spina:2024npx}. This research advances our understanding of gravitational physics and tests theoretical models against observational data \cite{Aoki:2024ktc,LISACosmologyWorkingGroup:2019mwx}. These recent findings can provide a solid foundation and context for work on scalar perturbations \cite{Aoki:2023bmz}.

In the study of cosmological scalar perturbations, Horndeski-like gravity offers a compelling framework that extends beyond the limitations of General Relativity \cite{Yu:2018qzl}. Horndeski gravity, being the most general scalar-tensor theory with second-order field equations \cite{Heisenberg:2018vsk,Horndeski:1974wa,Bruneton:2012zk}, provides a rich structure for exploring modifications to gravitational dynamics. This framework is particularly relevant for understanding the behavior of scalar perturbations in the early universe \cite{Ahmedov:2023num}, which play a crucial role in forming large-scale structures and the cosmic microwave background anisotropies \cite{Bernardo:2021qhu,Bahamonde:2021dqn}. By examining the scattering of scalar particles within this context, we can gain deeper insights into the stability and evolution of cosmological perturbations \cite{Jenks:2024fiu}. This research enhances our theoretical understanding and has potential implications for observational cosmology, offering new insight's to test the validity of Horndeski-like models against empirical data \cite{deBoe:2024gpf,Bahamonde:2021dqn}. 

In this work, we explore the ADM (Arnowitt-Deser-Misner) formalism applied to the scalar perturbations within the framework of Horndeski gravity \cite{Kobayashi:2019hrl}. Horndeski gravity, the most general scalar-tensor theory yielding second-order field equations, has garnered significant attention due to its ability to encompass a wide range of cosmological and astrophysical phenomena \cite{Bahamonde:2019ipm,Bahamonde:2019shr,Ezquiaga:2017ekz,Creminelli:2017sry,Baker:2017hug}. By employing the ADM formalism, we decompose the spacetime metric into a set of variables that facilitate the analysis of dynamical systems, particularly in the context of scalar perturbations \cite{Kobayashi:2019hrl,Santos:2021guj}. This approach not only enhances our understanding of the stability and evolution of perturbations in Horndeski models but also provides a robust framework for comparing theoretical predictions with observational data \cite{Ezquiaga:2017ekz,Creminelli:2017sry,Baker:2017hug}. Our work builds upon previous studies, such as those by \cite{Kobayashi:2019hrl,Santos:2021guj}, which have laid the groundwork for understanding perturbative dynamics in modified gravity theories. Through this analysis, we aim to contribute to the ongoing discourse on the viability of Horndeski gravity as a candidate for explaining dark energy.

%%%%%%%%%%%%%%%%%%%%%%%%%%%%%%%%%%%%%%%%%%%%%%%%%%%%%%%%%%%%%%%%%%%%%%%%%%%%%%%%%%%%%%%%%%%%%%%%%%%%%%%%%%%%%%%%

Following the previous works (e.g. \cite{Santos:2023flb, Santos:2024ynr}), we will consider the Horndeski-like Lagrangian
\begin{equation}\label{eq:Horn}
\mathcal{L}_H = \kappa(R-2\Lambda) - \frac{1}{2}(\alpha g_{\mu \nu} - \gamma G_{\mu \nu})\nabla^\mu \phi \nabla^\nu \phi,
\end{equation}
where $\kappa := (16\pi G_N)^{-1} = M_{\mathrm{pl}}^2/2$ and we consider an additional scalar field $\chi$ such that the action reads 
\begin{equation}\label{eq:action}
S = \int d^4 x \sqrt{|g|} (\mathcal{L}_H - \frac{1}{2}\nabla_\mu \chi \nabla^\mu \chi - V(\phi,\chi)).     
\end{equation}
In the remainder of the work we assume adopt the unit convention in which $\kappa = 1/4$. \\

To reproduce the standard Einstein Horndeski-like equations in the general-relativistic limit, we need to add the boundary term $S_\Sigma$ \cite{Santos:2023flb}:
\begin{equation}
S_{\Sigma} = 2\int d^3 x \sqrt{|g|} (\mathcal{L}_{\Sigma,H}+\mathcal{L}_{ct}),     
\end{equation}    
where 
\begin{align}
\mathcal{L}_{H,\Sigma}&= (K-\Sigma) - \frac{\gamma}{4}\nabla_\mu \phi \nabla_\nu K^{\mu \nu} \nonumber \\
&- \frac{\gamma}{4}(\nabla_\mu \phi \nabla_\nu \phi n^\mu n^\nu - (\nabla \phi)^2)K, \nonumber \\
\mathcal{L}_{ct} &= c_0 + c_1 R + c_2 R_{ij}^2 + c_3 R^2 + b_1 (\partial_i \phi \partial^i \phi)^2 + ... \nonumber \\
\end{align}
The full action includes matter is then $S_{total} = S + S_{\Sigma} + S_m$, where $S_m$ is the matter action. \\

The structure of the work will be as follows. In section \ref{sec:develop} we discuss the stability conditions of the scalar sector and compare it with that of the tensor sector under the assumption that $\delta \chi = 0$, which amounts to assuming that there is only one propagating scalar degree of freedom in the theory. Using the first-order formalism we will investigate analytical solutions in terms of a superpotential focusing on late times such that the matter sector is neglected in the analysis, which allow us to derive stability bounds on the allowed parameter space $(\alpha,\gamma)$. With the first-order formalism will show that there exist different background solutions, but for simplicity we will assume that $W=W(\phi)$ and $\chi=0$ in the remainder of the work. In section \ref{sec:slip} we will investigate the gravitational slip under the quasi-static approximation for different values of $k$ as a replacement of the shear viscosity that is used in the context of black holes. In section \ref{sec:particle} the entropy corresponding to particle production is computed, from which additional bounds follow that we can call entropy bounds. Other entropy bounds can be derived from considering the entropy of the apparent horizon, which will be done in section \ref{sec:app}. In section \ref{sec:app} we will also study the gravitational slip (minus one) to (apparent) entropy ratio. In section \ref{sec:conclusions} we provide the main conclusions, comment on the general philosophy and future prospects or directions that could follow from this work.   

\section{Propagation of scalar perturbations} \label{sec:develop}

Let $ds^2 = -N^2 dt^2 + \gamma_{ij}(dx^i + N^i dt)(dx^j + N^j dt)$, where $N=1+\tilde{\alpha}$ \footnote{We use the symbol $\tilde{\alpha}$ for the perturbation in the lapse, while the common notation is $\alpha$, however we do not want to introduce confusion with the coupling constant $\alpha$.}, $N_i = \partial_i \beta$ and $\gamma_{ij} = a^2 e^{2\zeta}(e^h)_{ij}$. We will focus on scalar perturbations so we will set $h_{ij}=0$ and work in the unitary gauge ($\delta \phi = 0$). We will also assume that $\chi$ is a background field so that $\delta \chi = 0$. When expanding the action the coefficients of $\zeta^2$ and $\tilde{\alpha}\zeta$ will vanish by the background equations and the $\tilde{\alpha}$,$\beta$ can be expressed in terms of $\zeta$ so that the quadratic action for scalars becomes \cite{Kobayashi:2011nu} \footnote{We note that the factor $e^{-\phi}/\lambda_s^3$ as in \cite{Santos:2024ynr} could be taken into account upon defining the energy-momentum tensor $T_{\mu \nu}^{(m)}$. The form of the field equation in \cite{Santos:2021guj} is equivalent under redefinition of the energy-momentum tensor.}:
\begin{eqnarray}
S^{(2)}_S= \int dt d^3 x\,a^3 \left[
{\cal G}_S
\dot\zeta^2
-\frac{{\cal F}_S}{a^2}
(\Vec{\nabla}\zeta)^2
\right]\label{scalar2},
\end{eqnarray}
where
\begin{eqnarray}\label{eq:stab}
{\cal F}_S&:=&\frac{1}{a}\frac{d}{dt}\left(\frac{a}{\Theta}{\cal G}_T^2\right)
-{\cal F}_T>0,
\\
{\cal G}_S&:=&\frac{\Sigma }{\Theta^2}{\cal G}_T^2+3{\cal G}_T>0,
\end{eqnarray}
and the scalar sound speed squared is $c_S^2:= \mathcal{F}_S/\mathcal{G}_S$. For the action under consideration we have that the part of the action without $\chi$ in terms of Horndeski functions takes the form \cite{Kobayashi:2011nu} \footnote{Note that here we have absorbed the $\kappa$ into $\alpha,\gamma$ compared to \cite{Santos:2023flb}.}:
\begin{align}
K(\phi,X) &= -\frac{1}{2}\Lambda + \alpha X \nonumber \\
G_3(\phi,X) &= 0 \nonumber \\
G_4(\phi,X) &= \frac{1}{4} \nonumber \\
G_5(\phi,X) &= -\frac{\gamma}{2}\phi, 
\end{align}
where $X := -\frac{1}{2}g_{\mu \nu}\nabla^\mu \phi \nabla^\nu \phi$. Since the terms with $\chi$ do not contribute to the tensor sector we directly conclude: $\mathcal{G}_T = 2(G_4 + X G_{5\phi}) = (1-\gamma \dot{\phi}^2)/2$ and $\mathcal{F}_T = 2(G_4 - X G_{5 \phi}) = (1 + \gamma \dot{\phi}^2)/2$ since $X=\dot{\phi}^2/2$. This gave in the previous work \cite{Santos:2024ynr} precisely $c_{\mathrm{GW}}^2 = \mathcal{F}_T/\mathcal{G}_T = (1 + \gamma \dot{\phi}^2)/(1 - \gamma \dot{\phi}^2)$. The stability conditions for tensors are $\mathcal{F}_T,\mathcal{G}_T>0$ \cite{Kobayashi:2011nu}. The conditions become $\gamma \dot{\phi}^2<1$ (see also \cite{Santos:2021guj}). However, it is also known that GW170817 and GRB 170817A contrain the allowed value of $-3\cdot 10^{-15} \leq c_{\mathrm{GW}}-1 \leq 7 \cdot 10^{-16}$ for redshift $z \leq 0.009$ \cite{Kase:2018aps}. This has also consequences for the allowed $\gamma$ and $\phi(t)$: $1 \gg |\gamma \dot{\phi}^2(t_0)|$. For general $\phi(t)$ of course this puts a stringent constraint on the allowed values of $\gamma$. The argument however is $k$-dependent as argued in \cite{deRham:2018red} so in general one could have $c_T = c_T(k)$ for the speed of tensors. The stability conditions and entropy bounds discussed in this work should be treated as separate bounds from gravitational wave speed constraint. Despite the fact that the gravitational wave speed constraint is very stringent compared to the other bounds, the goal of this work is to illustrate the broader philosophy. For more general subclasses of Horndeski theory the stability conditions together with the entropy bounds can provide constraints independently of the gravitational wave speed constraint. The specific choice of the model is due to the fact that the model admits analytical background solutions and that the entropy of the apparent horizon can be computed analytically. This allows us to systematically compare the stability conditions and the entropy bounds. Next, we will follow the procedure described in \cite{Santos:2024ynr} we can try to find background solutions by determining possible superpotentials $W(\phi,\chi)$. This allows then to find $\phi(t),\chi(t),H(t)$ and $a(t)$ describing the background evolution of the system. Let us define the superpotential through the following equations using the first-order formalism:
\begingroup
\setlength{\abovedisplayskip}{5pt} 
\setlength{\belowdisplayskip}{5pt}
\begin{align}\label{eq:background}
\dot{\phi} &= -W_{\phi} \nonumber \\
\dot{\chi} &= -W_{\chi} \nonumber \\
H &= W.
\end{align}
\endgroup
where $W_\phi = \partial W/\partial \phi$ and $W_\chi = \partial W/\partial \chi$. The goal of the introduction of the superpotential is to reduce the second order differential equations to first order, which in general would be easier to solve. Let us investigate the stability of the scalar sector of the action. For the scalar sector the part of the action with $\chi$ does contribute. It amounts to the replacement $XK_X \mapsto XK_X + \frac{\dot{\chi}^2}{2}$ in the expression for $\Sigma$ \cite{Kobayashi:2011nu}. We then have that $\Sigma = X K_X -6H^2 (G_4 + 6X G_{5\phi}) + \frac{1}{2}\dot{\chi}^2 = \frac{\alpha \dot{\phi}^2}{2} -\frac{3}{2}H^2 (1 - 6\gamma \dot{\phi}^2) + \frac{1}{2}\dot{\chi}^2$ and $\Theta = 2H(G_4 + 3XG_{5\phi}) = \frac{1}{2}H(1 - 3\gamma\dot{\phi}^2)$. 
Given these expressions we can compute $\mathcal{F}_S$ and $\mathcal{G}_S$ defined in equation (\ref{eq:stab}):
\setlength{\abovedisplayskip}{5pt} 
\begin{align}\label{eq:stab1}
\mathcal{F}_S &= \frac{(1 - \gamma \dot{\phi}^2)^2}{2(1 - 3\gamma \dot{\phi}^2)} - \frac{2(1 - \gamma \dot{\phi}^2) \gamma \dot{\phi}\ddot{\phi}}{H(1 - 3\gamma \dot{\phi}^2)}- \frac{1}{2} - \frac{1}{2}\gamma \dot{\phi}^2 \nonumber \\
&-\frac{(1 - \gamma \dot{\phi}^2)^2}{2H^2(1-3\gamma \dot{\phi}^2)^2}[\dot{H}(1 - 3\gamma \dot{\phi}^2) - 6 H\gamma \dot{\phi}\ddot{\phi}] \nonumber \\
&= \frac{(1 - \gamma W_\phi^2)^2}{2(1 - 3\gamma W_\phi^2)} + \frac{2(1 - \gamma W_\phi^2)\gamma W_\phi^2 W_{\phi \phi}}{W(1 - 3\gamma W_\phi^2)} \nonumber \\
&-\frac{(1 - \gamma W_\phi^2)^2}{2W^2(1 - 3\gamma W_\phi^2)^2}[-(W_\phi^2 + W_\chi^2)(1 - 3\gamma W_\phi^2)\nonumber \\ 
&+6\gamma WW_\phi^2 W_{\phi \phi}]-\frac{1}{2}-\gamma W_\phi^2/2, 
\end{align}
\begin{align}\label{eq:stab2}
\mathcal{G}_S &= \frac{-\frac{3}{2}H^2(1 - 6\gamma \dot{\phi}^2) + \frac{1}{2}\dot{\chi}^2 + \frac{\alpha}{2} \dot{\phi}^2}{H^2(1 - 3\gamma \dot{\phi}^2)^2}(1 - \gamma \dot{\phi}^2)^2 \nonumber \\
&+ \frac{3}{2} - \frac{3}{2}\gamma \dot{\phi}^2 \nonumber \\
&= \frac{-\frac{3}{2}W^2(1 - 6\gamma W_\phi^2) + \frac{1}{2}W_\chi^2 + \frac{\alpha}{2}W_\phi^2}{W^2(1 - 3\gamma W_\phi^2)^2}(1 - \gamma W_\phi^2)^2 \nonumber \\
&+\frac{3}{2} - \frac{3}{2}\gamma W_\phi^2. 
\end{align}

Given a field $\phi(t)$ and $H(t)$ that satisfy the background equations, it is possible to check whether the the no-ghost and no gradient instability conditions $\mathcal{F}_S>0$ and $\mathcal{G}_S>0$ are satisfied at all times during the cosmic history \footnote{Obviously with all times we mean all times at which the matter/radiation contributions can be neglected compared to the contribution of the scalar field $\phi$.}. The background field equations for the action (\ref{eq:action}) are found to be \cite{Kobayashi:2011nu}:
\begin{eqnarray}
H^2 &=& \frac{\Lambda + \alpha \dot{\phi}^2 + \dot{\chi}^2 + 2V(\phi,\chi)}{3(1-3\gamma \dot{\phi}^2)} \nonumber \\
-\frac{1}{2}(3H^2 + 2\dot{H}) &=& -\frac{1}{2}\Lambda - V(\phi,\chi) + \frac{\alpha}{2} \dot{\phi}^2 + \frac{1}{2}\dot{\chi}^2 \nonumber \\
&-&\gamma[\dot{H}\dot{\phi}^2 + 2H\dot{\phi}\ddot{\phi} + 3H^2\dot{\phi}^2/2].  
\end{eqnarray}
These background equations are only valid at late times for which the matter and radiation contributions are neglected in the expressions \footnote{There exist methods to extend the first-order formalism to include also matter contributions, see e.g. \cite{Bazeia:2006mh}.}. The reason for this prescription is that it is easier to find analytical solutions to the background equations. The equations of motion of the two scalar fields $\phi$ and $\chi$ can also be found \cite{Kobayashi:2011nu}:
\begin{eqnarray}
\ddot{\phi} + 3H\dot{\phi} + \frac{6\gamma H\dot{H}\dot{\phi}}{\alpha + 3\gamma H^2} + \frac{1}{\alpha + 3\gamma H^2}\frac{\partial V}{\partial \phi} &=& 0 \nonumber \\
\ddot{\chi} + 3H \dot{\chi} + \frac{\partial V}{\partial \chi} &=& 0. 
\end{eqnarray}
From combining the Friedmann equations and the combination of the first Friedmann equation with the scalar field equation for $\phi(t)$ respectively it follows that
\begin{eqnarray}
0 &=& WW_{\phi \phi} + \frac{3}{2}W^2 - \frac{1}{2\gamma}(1 - \alpha) + \frac{1}{2}W_\phi^2 + \frac{1}{2}W_\chi^2, \nonumber \\   
0 &=& WW_{\phi \phi} + \frac{3}{2}W^2 - \frac{1}{2\gamma}(1-\alpha) + \frac{1}{2}W_\phi^2 \nonumber \\
&-& W_\chi^2 + \frac{1}{6\gamma} \frac{W_{\phi \chi}W_\chi}{WW_\phi}.  
\end{eqnarray}

\begin{figure}
    \centering
    \includegraphics[width=0.9\linewidth]{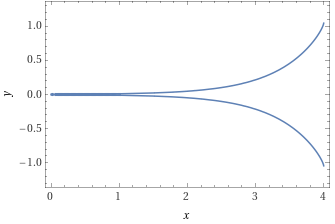}
    \caption{On the $y$-axis is related to the scalar field $\phi$ and on the $x$-axis is the time $t$ in some units (here we have chosen $\tilde{c}_1 = -4$ such that $\phi(0)=0$). The plot assumes the superpotential $W(\phi) = \cos^{2/3}(3\phi/2)$. Here the two branches of the solution are illustrated.}
    \label{fig:scalar}
\end{figure}

The equations reduce to equation (12) in ref. \cite{Santos:2019ljs} when the $\chi$ field is not considered. From the two equations we obtain the following relation: 
\begin{equation}\label{eq:superp}
9\gamma W_\chi W_\phi = \frac{W_{\phi \chi}}{W}.    
\end{equation}
The result is however not in agreement with the previous work \cite{Santos:2024ynr} in which it was proposed that there was an exact mapping between the model and the corresponding braneworld scenario. This implies that the search for analytical solutions for this model is more complicated. A trivial solution of equation (\ref{eq:superp}) would be the case $W=\mathrm{constant}\neq 0$, which would correspond to $H = H_0$. This means that there exists a solution of the type $a(t)\propto e^{H_0(t-t_0)}$, which can describe the cosmic acceleration at late times. Also the solutions $W = W(\phi)$ and $W = W(\chi)$ satisfy equation (\ref{eq:superp}). In general however it is complicated to find analytically cosmological solutions from equation (\ref{eq:superp}). Set let us focus on solutions with $W = W(\phi)$ and set $\chi = 0$ for simplicity in the remainder of the work following the approach of \cite{Santos:2019ljs}. The extension with the inclusion of an additional field $\chi$ we leave for a future work. Define the parameter $\beta := \frac{1}{\gamma}(1-\alpha)$, then the equation for the superpotential becomes:
\begin{equation}\label{eq:superphi}
2WW_{\phi \phi} + W_\phi^2 + 3W^2 - \beta = 0.    
\end{equation}
For the case that $\beta = 0$ there exist analytical solutions, however if $\beta \neq 0$ then one needs to resort to numerical methods to solve for the background equations \cite{Santos:2019ljs}. The equations (\ref{eq:background}) and (\ref{eq:superphi}) can be solved numerically under the boundary conditions $W(0) = W_\phi(0)=1$ and $\phi(0) = 0$, where $t=0$ does not describe the Big-Bang but refers to the same at which the formalism starts to describe the present phase of the Universe. The exploration of the model at earlier times in the cosmic history we leave for a future work as the goal is to describe the late-time cosmology. This can be done using for instance \texttt{EFTCAMB} \footnote{\url{https://eftcamb.org/}}. The advantage of using the first-order approach as prescribed is that we do not need to specify explicitly the potential $V(\phi)$ and that it still allows to find solutions that predict a de Sitter phase at late-time cosmology \cite{Santos:2019ljs}. Focusing on the dark energy sector of the theory we find for the dark energy density and pressure \cite{Kobayashi:2011nu} \footnote{We define the effective energy density and pressure of dark energy via: $\rho := \frac{3}{2}H^2$ and $p := -\frac{1}{2}(3H^2 + 2\dot{H})$.}:
\begin{eqnarray}\label{eq:rho-p}
&&\rho = \frac{1}{2}\Lambda + V(\phi) + \frac{9\gamma}{2}W_\phi^2 W^2 + \frac{\alpha}{2}W_\phi^2  \nonumber \\
&&p =-\frac{1}{2}\Lambda -V(\phi) + \frac{\alpha}{2}W_\phi^2  \nonumber \\
&&-\gamma[-W_\phi^4 - 2 WW_\phi^2 W_{\phi \phi} + 3W^2 W_\phi^2/2].
\end{eqnarray}      
As a consistency check note that indeed if the scalar field equation is satisfied then $-V_\phi = (\alpha + 3\gamma W^2)W_\phi W_{\phi \phi} -3WW_\phi(\alpha + 3\gamma W^2) + 6\gamma WW_\phi^3$, which implies that $\dot{\rho}+3H(\rho+p) = 0$ as expected. The search for analytical solutions can be simplified by assuming $\alpha \geq 0$ and performing a field redefinition $\phi \mapsto \sqrt{\alpha}\phi =: \varphi$, which transforms the kinetic term into a canonical one. Note that this transforms the action such that $\gamma \mapsto \gamma/\alpha$. Since the field redefinition just amounts to a redefinition of coupling constants, we will write $\phi$ for the canonical scalar field as well. The Horndeski functions for the canonical scalar field are then $K(\phi,X) = -\frac{1}{2}\Lambda + X - V(\phi/\sqrt{\alpha})$, $G_3 = 0$, $G_4 = 1/4$ and $G_5 = -\frac{\gamma}{2\alpha}\phi$. The equation for the superpotential becomes then $2WW_{\phi \phi} + W_\phi^2 + 3W^2 = 0$ \footnote{In the remainder of the work we will adopt a redefinition of coupling constants via $\alpha \frac{\dot{\phi}^2}{2} \mapsto \frac{1}{2}\dot{\phi}^2$ and $\gamma \mapsto \gamma/\alpha$ assuming $\alpha \geq 0$.}. This has the analytical solution $W(\phi) = c_1 \cos^{2/3}(c_2 + 3\phi/2)$, where $c_1,c_2$ are some constants. Let us focus on the solution with $c_1 = 1$ and $c_2 = 0$ we checked that expanding this equation about $\phi \approx 1$ at $t = 0$ indeed gives equation (14) in ref. \cite{Santos:2019ljs} describing the cosmic acceleration at late times. However, this approximation relies on a Taylor series and to circumvent it we can solve the set of differential equations more exactly. From the equation $\dot{\phi} = -W_\phi$ we find that the following equation needs to be satisfied:
\begin{equation}
\tilde{c}_1 + t = -\frac{1}{2} \cos^{4/3}(3\phi/2) \ _2F_1(2/3, 1, 5/3, \cos^2(3\phi/2)),    
\end{equation}
where $\tilde{c}_1$ is some constant and $_2F_1$ denotes the hypergeometric function. $\phi(t)$ can be found by making an implicit plot (see Figure (\ref{fig:scalar})) and the result is that we can set $\phi \in [0,1]$ \footnote{There is in fact also the solution $\phi \mapsto -\phi$, which we do not consider since $W(\phi) = W(-\phi)$ if $c_2=0$.} since $\phi$ increases monotonically  starting at some time $t^\star$ at which $\phi(t^\star) = 0$ and is zero at earlier times $t<t^\star$. In the superpotential solution $W(\phi) = \cos^{2/3}(3\phi/2)$ we will therefore use the approximation that $\phi \in [0,1]$ for the times at which the formalism approximates the cosmology. The dark energy equation-of-state could be computed via $\omega = p/\rho = -1+\frac{2}{3}\frac{W_\phi^2}{W^2}$. The result is that the solution satisfies $w<-1/3$ only for $0 \leq \phi \lessapprox 0.5$. The regime with $\phi>0.5$ cannot describe the cosmic acceleration and thus does not describe the late Universe. Therefore in the following we will have to assume that this solution cannot describe the ultimate cosmic evolution of the Universe but rather some phase of the cosmic history in which cosmic acceleration occurs. Note also that $a = \exp(-\int \frac{W}{W_\phi}d\phi) = \exp(\cot(3\phi/2))$ for the superpotential under consideration. This means that formally the scale factor is divergent for $\phi \rightarrow 0$, which can be circumvented by noting that $w \approx -1$ holds for $\phi \lessapprox 0.1$, hence during those times we may approximate the model by just the usual LCDM. Therefore in this work we will focus on times in the cosmic evolution with $\phi \in [0.1,0.5]$. For the stability conditions of the scalar sector of the theory we find for this choice that $\gamma \lessapprox 1 + 1.25\alpha$ (where also $\alpha \geq 0$). The tensor constraint $\gamma \dot{\phi}^2<1$ becomes just $\frac{\gamma}{\alpha}W_\phi^2 < 1$ after canonical normalization. This implies $\gamma/\alpha \lessapprox 1.75$ since $W_{\phi}^2$ is monotonically increasing and $\mathrm{max}(W_{\phi}^2)\approx 0.57$. Therefore the tensor stability constraint is stronger than the scalar stability constraint, but this result of course depends on this choice of the superpotential. All relevant quantities are expressed in terms of the superpotential and can therefore be computed easily in this fashion.     

%%%%%%%%%%%%%%%%%%%%%%%%%%%%%%%%%%%%%%%%%%%%%%%%%%%%%%%%%%%%%%%%%%%%%
\section{Perturbation theory and gravitational slip}\label{sec:slip}
%%%%%%%%%%%%%%%%%%%%%%%%%%%%%%%%%%%%%%%%%%%%%%%%%%%%%%%%%%%%%%%%%%%%%

Utilizing the Newtonian gauge, we employ the following metric to facilitate our analysis \cite{Pogosian:2016pwr}:
\begin{eqnarray}
ds^2&=&g_{\mu\nu}dx^{\mu}dx^{\nu}\label{metr}\\
&=& -(1+2\Psi)dt^2+a^2(t)(1-2\Phi)\delta_{ij}dx^idx^j,\nonumber
\end{eqnarray}
where $\Psi(t,x^i),\Phi(t,x^i)$ are perturbations. We will work in the linear regime of the theory. The shear viscosity was studied in \cite{Barbosa:2018iiq} in the context of General Relativity. The shear viscosity in that case is proportional to the difference in $\Psi$ and $\Phi$. In the case when anisotropic stress is negligible, in modified gravity the difference between $\Psi$ and $\Phi$ can still be non-zero and this is described by the so-called gravitational slip \cite{Pogosian:2016pwr}. We will following along these lines not take into an explicit anisotropic matter contribution but the inclusion is possible \cite{Zucca:2019xhg}. In the previous work \cite{Feng:2015oea} the shear viscosity to entropy ratio for black holes was considered for the Horndeski theory in this work and provided the bound $\eta/S \geq 1/4\pi$. In the context of the AdS/BCFT correspondence applied to black hole thermodynamics \cite{Santos:2023flb,Santos:2023mee}, scalar fields break conformal symmetry, leading to a notable increase in entropy that diminishes with rising temperature. This behavior parallels conformal time. An increasing $\eta/S$ ratio may be linked to the fluid dynamics of the early universe, offering insights into the fundamental physics of cosmic evolution. During inflation or reheating, scalar fields can decay into other particles, contributing to the universe's entropy. In this decay process, the information captured by $\eta/S$ decreases, necessitating other transport coefficients to capture the information of newly created particles. Interactions and decay processes of scalar particles can significantly alter the ratio $\eta/S$, where $\eta$ is the shear viscosity and 
$S$ is the entropy density. Efficient creation and scattering of scalar particles can increase the entropy density \cite{Yu:2018qzl,Dai:2021dog}, potentially reducing the $\eta/S$ ratio. However, for the action considered in this work the notion of shear viscosity is non-trivial in the cosmological context since the anisotropic stress and shear tensor are not simply related by the usual Eckart's constitutive relations \cite{Giusti:2021sku}. This makes the computation of the shear viscosity complicated for the Horndeski-like action in this work. This serves as a motivation to replace the usual shear viscosity to entropy ratio by the gravitational slip (minus one) to entropy ratio, where the entropy will be the one of the apparent horizon. The question is thus whether we can gain any physical insights by studying the gravitational slip to entropy ratio. For this we compute the gravitational slip for the action (\ref{eq:action}) in the large scale and small scale limit \cite{Pogosian:2016pwr}. From the equations in \cite{Pogosian:2016pwr} and \cite{Bellini:2014fua} \footnote{Note that the work \cite{Pogosian:2016pwr} contains a typo in the expression for $\alpha_K$, which is correct in the original paper \cite{Bellini:2014fua}.} we find that for the action (\ref{eq:action}) in terms of the superpotential $W$ \footnote{Not applying the canonical normalization yet, also in equations for gravitational slip below.}:  
\begin{eqnarray}
M_\star^2 &=& \frac{1}{2}(1 - \gamma W_\phi^2) \nonumber \\ 
\alpha_B &=& \frac{4\gamma W_\phi^2}{1 - \gamma W_\phi^2} = 2 \alpha_T \nonumber \\
\alpha_M &=& \frac{2\gamma W_\phi^2 W_{\phi \phi}}{W(1 - \gamma W_\phi^2)}\nonumber \\
\alpha_K &=& \frac{4W_\phi^2(\alpha + 3W^2 \gamma)}{W^2(1 - \gamma W_\phi^2)^2}.
\end{eqnarray}
This allows us to compute the gravitational slip \footnote{When referring to the gravitational slip in general we use the symbol $\tilde{\gamma}$ to avoid confusion with the coupling constant $\gamma$.} in the limits $k/a \ll M$ and $k/a \gg M$, called $\gamma_0$ and $\gamma_\infty$ respectively \cite{Pogosian:2016pwr}:
\begin{equation}
\gamma_0 = \frac{1 - \gamma W_\phi^2}{1 + \gamma W_\phi^2}, 
\end{equation}
\begin{equation}
\gamma_\infty = \frac{(1 - \gamma W_\phi^2)f + \gamma W_\phi^2\Big[\frac{2\gamma^2 W_\phi^4}{1-\gamma W_\phi^2} + \frac{\gamma W_\phi^2 W_{\phi \phi}}{W}\Big]}{(1 - \gamma W_\phi^2)f + 2\gamma W_\phi^2 f + \Big[\frac{2\gamma^2 W_\phi^4}{1-\gamma W_\phi^2} + \frac{\gamma W_\phi^2 W_{\phi \phi}}{W}\Big]^2},    
\end{equation}
where $f$ is defined as
\begin{eqnarray}
f &=& -\frac{\gamma W_\phi^2 W_{\phi \phi}}{2W} - \frac{2\gamma^2 W_\phi^4 W_{\phi \phi}}{W(1 - \gamma W_\phi^2)} + \frac{\gamma^2 W_\phi^4}{(1-\gamma W_\phi^2)} \nonumber \\
&-& \frac{\gamma^3 W_\phi^6}{2(1 - \gamma W_\phi^2)} - \frac{W_\phi^2}{4W^2} \Big(-1 + 3\gamma W_\phi^2 \Big),  
\end{eqnarray}
where we did not include $\Omega_{m,0}$ in the expression according to the previously discussed assumption in the first-order formalism. Given a background cosmology prescribed by $\phi(t)$ and $H(t)$ it is possible to compute $\gamma_0,\gamma_\infty$ analytically using these expressions. The equations rely on the so-called quasi-static approximation (QSA), which is in general model dependent and should in principle be checked for each model separately \cite{Pogosian:2016pwr}. The QSA can be explicitly checked with the necessary but not sufficient condition that the given Fourier mode lies within the scalar field's sound horizon at all times under consideration, i.e. $k/(aH)>c_s$ for all $a$, where $c_s$ is the speed of sound of the scalar perturbations \cite{Pogosian:2016pwr,Peirone:2017ywi}. In \cite{Peirone:2017ywi} it was shown in the sampling of full Horndeski theory that for $k>0.001h \ \mathrm{Mpc}^{-1}$ approximately only one percent fails to satisfy the QSA. Note that the condition $k/(aH) > c_s$ in terms of the superpotential and the scalar field becomes:
\begin{eqnarray}
k &>& W\exp(-\int \frac{W}{W_\phi}d\phi) \Big[\frac{2W^2(1 - \gamma W_\phi^2)}{W_\phi^2 (\alpha + 3\gamma W^2) + 6\gamma W_\phi^4 W^2 }\nonumber \\
&\Big[&-\frac{\gamma W_\phi^2 W_{\phi \phi}}{2W} - \frac{2\gamma^2 W_\phi^4 W_{\phi \phi}}{W(1 - \gamma W_\phi^2)} + \frac{\gamma^2 W_\phi^4}{1-\gamma W_\phi^2} \nonumber \\
&-& \frac{\gamma^3 W_\phi^6}{2(1 - \gamma W_\phi^2)} - \frac{W_\phi^2}{4W^2} \Big(-1 + 3\gamma W_\phi^2\Big)\Big]\Big]^{1/2}. 
\end{eqnarray}
In the Figures \ref{fig:smallk-ratio} and \ref{fig:largek-ratio} we will implicitly assume values of $k$ such that this condition is satisfied. Note that this in general may depend on the model parameters $(\alpha,\gamma)$. An approach that would not rely on the QSA would be that of for instance \texttt{EFTCAMB} in which case the cosmological perturbations are evolved from radiation era $a \sim 10^{-8}$ up to today $a=1$. According to the first-order formalism approach we adopt in this work by which we focus on analytical solutions, we leave the exploration beyond the QSA, the comparison to the QSA and thus also the check of its validity for a future work.

%%%%%%%%%%%%%%%%%%%%%%%%%%%%%%%%%%%%%%%%%%%%%%%%%%%%%%%%%%%%%%%%%%%%%%%%
\section{Particle production}\label{sec:particle}
%%%%%%%%%%%%%%%%%%%%%%%%%%%%%%%%%%%%%%%%%%%%%%%%%%%%%%%%%%%%%%%%%%%%%%%%

We now turn our attention to the entropy associated with scalar perturbations, as discussed in previous studies \cite{Yu:2018qzl,Prigogine:1988jax,Kaur:2021dix,Dai:2021dog,Cipriano:2023yhv}. In this analysis, we will disregard the contributions to entropy from spacetime and the apparent horizon \cite{Yu:2018qzl,Prigogine:1988jax}, thus focusing solely on the entropy relevant to scalar particles. The entropy of the apparent horizon we will address in the next section, while the entropy from spacetime we will not investigate in this work. Under these assumptions, the total entropy of the universe can be expressed in a differential form, comprising two distinct components. This approach allows us to isolate the effects of scalar perturbations on the entropy dynamics, providing a clearer understanding of their role in the cosmological context.

We can express the total entropy consisting of two parts:
\begin{eqnarray}
dS_{in}=dS_f+dS_c\label{Ent}.
\end{eqnarray}
In this equation $dS_f$ represents the entropy flux and $dS_c$ the entropy production. For stable thermodynamic systems without particle production, as long as there is no energy exchange with the outside, we will have $dS_f = 0$ and $dS_c=0$, and therefore $dS_{in}=0$. However, for our system in question, which can be considered an isolated system, only $dS_f$ vanishes. In this case, the total entropy can be expressed as \cite{Yu:2018qzl,Prigogine:1988jax} where the specific entropy of the scalar particle is constant. Under this assumption and neglecting interactions with other particles than $\phi$, the scalar particle production rate $\Gamma_\phi$ for the action can be calculated. 
The particle production rate $\Gamma_\phi$ is then found by \cite{Yu:2018qzl,Prigogine:1988jax}: 
\begin{eqnarray}
\Gamma_\phi := \frac{\dot{\rho}_\phi + 3H(\rho_\phi+p_\phi)}{\rho_\phi+p_\phi},    
\end{eqnarray}
where $\rho_\phi := \frac{\alpha}{2}W_\phi^2 + V(\phi)$ and $p_\phi := \frac{\alpha}{2}W_\phi^2 - V(\phi)$. The $\rho_\phi,p_\phi$ can be computed from equation (\ref{eq:rho-p}). Using that $\dot{\rho}+3H(\rho+p) = 0$ it easily follows that the particle production rate is:
\begin{equation}
\Gamma_\phi = -\frac{9\gamma}{\alpha}W^3 + \frac{6\gamma}{\alpha}WW_\phi^2 + \frac{3\gamma}{\alpha}W^2 W_{\phi \phi}.    
\end{equation}
Under the canonical normalization of the scalar field, the expression for $\Gamma_\phi$ does not change upon redefinition of $\rho_\phi,p_\phi$. The caveat in the approach of \cite{Yu:2018qzl} is that the matter action is neglected in the computation of $\Gamma_\phi$. In a more precise approach, the matter should be included appropriately, which will alter the particle production rate. The entropy $S_{in}$ satisfies the equation $\frac{dS_{in}}{dt} = \Gamma_\phi S_{in}$. Given the evolution of $\Gamma_\phi$ with one can study for given background field $\phi(t)$ the evolution of entropy over time and require the bound $\frac{dS_{in}}{dt} \geq 0$ from the second law of thermodynamics \cite{Prigogine:1988jax}. Note that in terms of the superpotential: $-W_\phi \frac{dS_{in}}{d\phi} = \Gamma_\phi S_{in} \geq 0$. This gives the equation:
\begin{equation}
S_{in}(\phi) = S_{in}(\phi_0) \exp\Big(-\int \frac{\Gamma_\phi}{W_\phi}d\phi \Big),    
\end{equation}
where we assume that $S_{in}(\phi_0)>0$ so this gives the following bound
\begin{equation}
\Gamma_\phi \exp\Big(-\int \frac{\Gamma_\phi}{W_\phi}d\phi \Big) \geq 0.    
\end{equation}
Since the expression will be a function of $(\alpha,\gamma)$ it can be expected that this will constrain the allowed values for these parameters. We see that $\exp(-\int \frac{\Gamma_\phi}{W_\phi}d\phi) \propto \exp(-\gamma/\alpha)$, which is nonnegative, and hence the bound on the coupling constants translates into just $\Gamma_\phi \geq 0$, which is interpreted as particle production. We can check for which values of $\gamma/\alpha$ for a given superpotential $W$ the bound is respected. For the case of the superpotential $W(\phi) = \cos^{2/3}(3\phi/2)$. Since then $\Gamma_\phi = -\frac{9\gamma}{2\alpha}(1+2\cos(3\phi))$, we obtain the following entropy bound: $-\frac{\gamma}{\alpha}\Big(\frac{1}{2} + \cos(3\phi)\Big) \geq 0$. It follows that $\gamma \leq 0$ due to $\alpha \geq 0$ and $\frac{1}{2}+\cos(3\phi) \geq 0$ for all $\phi \in [0.1,0.5]$. This is consistent with the tensor and scalar stability constraint.

%%%%%%%%%%%%%%%%%%%%%%%%%%%%%%%%%%%%%%%%%%%%%%%%%%%%%%%%%%
\section{Entropy of the apparent horizon}\label{sec:app}
%%%%%%%%%%%%%%%%%%%%%%%%%%%%%%%%%%%%%%%%%%%%%%%%%%%%%%%%%%

In this section, we explore the entropy associated with the apparent horizon of the universe, a concept intrinsically linked to the radius of the apparent horizon \cite{Mimoso:2013zhp,Pavon:2012qn,Easson:2010av,Brustein:2007jj,Bak:1999hd}. The entropy of the apparent horizon plays a crucial role in understanding the thermodynamic properties of the universe, as it provides insights into the interplay between gravitational dynamics and thermodynamics. By examining this relation, we aim to shed light on the fundamental processes governing the evolution of the universe. Following the prescriptions of \cite{Santos:2021orr,Santos:2023flb,Santos:2023eqp,Santos:2023mee,Santos:2024qpv,Santos:2024zoh}, we have for the entropy of the apparent horizon:
\begin{equation}
S=\frac{1}{H^2}\left(1-\frac{\xi}{4}\right) ;\ \ \xi=\frac{\alpha+\gamma\Lambda}{\alpha}
\end{equation}
Here in our discussions, the radius of the apparent horizon is $r_h= H^{-1}$ for the FRW universe \cite{Bak:1999hd}. In this equation, we focus on the bulk contribution to the entropy. In the context and convention $8\pi G_N = 1$ of ref. \cite{Yu:2018qzl} we see that this translates into: 
\begin{equation}
G_{eff} = \frac{\pi}{2\Big(1-\frac{\xi}{4}\Big)}.    
\end{equation}
As commented on in \cite{Yu:2018qzl} the sum of the entropy $S_{in}$ and that of the apparent horizon $S_h \equiv S$, denoted $S_{sum} = S_{in} + S_h$, should be considered as well. In the previous section, the condition for $\dot{S}_{in}(t) \geq 0$ was derived. Let us consider the impact of the entropy bound $\dot{S}(t) \geq 0$ on the allowed parameter space, following the lines of \cite{Yu:2018qzl}. In case of the radiation or matter era it typically holds that $a(t)\propto t^n$, for which the constraint $\dot{S}(t) \geq 0$ translates into $1 \geq \xi/4$ or equivalently $3\alpha \geq \gamma \Lambda$. At late times close to $t \approx t_0$ we found that there exists a solution $a(t)\propto e^{H_0(t-t_0)}$ and this yields a trivial bound since $S$ is constant. For the superpotential $W(\phi) = \cos^{2/3}(3\phi/2)$ we observe that it holds that $\frac{d}{dt}(\frac{1}{H^2}) = -\frac{\dot{H}}{H^3} = \frac{W_\phi^2}{W^3} \geq 0$ for all $\phi \in [0.1,0.5]$ and thus we find again the bound $3\alpha \geq \gamma \Lambda$ \footnote{It is easily seen that the canonical normalization does not alter the entropy bound.}. Note that if $\gamma = 0$ it follows that $\alpha \geq 0$ must hold. The goal of this discussion is to illustrate that the requirement of that apparent entropy with time has a big impact on the allowed parameter space. In particular, note that for instance $\Lambda$ can be constrained with this bound, while this parameter is not explicitly present in the scalar and tensor stability constraints. The form of the entropy of the apparent horizon is different with respect to the discussion in \cite{Yu:2018qzl} for which $\dot{S}(t)\geq 0$ was trivially true for the cases we considered. These assumptions can be lightened by assuming that only $\dot{S}_{sum}(t) \geq 0$ hold and the intermediate times should be considered as well. The entropy bound on $S_{sum}$ is:
\begin{equation}
\Gamma_\phi S_{in}(\phi_0) \exp\Big(-\int \frac{\Gamma_\phi}{W_\phi}d\phi \Big) + \Big(1-\frac{\xi}{4}\Big)\frac{W_\phi^2}{W^3} \geq 0.     
\end{equation}
The exact bound however depends on the value of $S_{in}(\phi_0)$. This means that it is not simple to constrain the parameter space of coupling constants using $S_{sum}$. Furthermore, this expression does not take into account the entropy contribution from spacetime itself and it neglects the interaction of $\phi$ with the matter sector. In a future work, a robust development of entropy bounds for modified gravity theories should take all the caveats into account. \\

The impact of the study of $\eta/S$ for black holes gave interesting insights and the same is expected for cosmology, however as discussed the shear viscosity for the Horndeski theory in this work is complicated to define. Because of this we propose to replace it by the study of the gravitational slip (minus one) to entropy ratio $(\tilde{\gamma}-1)/S$. Of course measurements galaxy clusters can efficiently constrain the gravitational slip, however the goal of our work is to propose that its ratio with the entropy could be a replacement for the usual shear viscosity to entropy ratio for black holes. \\

We obtain the following Figures \ref{fig:smallk-ratio} and \ref{fig:largek-ratio}. It can be seen that for $\phi \lessapprox 0.1$ it follows that to good approximation $\tilde{\gamma} \approx 1$ confirming that the solution behaves like LCDM in this regime. From the figures we also note that once $\gamma \rightarrow 0$ that then the gravitational slip goes to unity (the value in LCDM). In the figures we checked for the scalar stability and the entropy bound from the apparent horizon. For $\gamma \gtrapprox 0.01$ one can notice, independently of the value of $k$, a deviation from the LCDM value of the gravitational slip at late times. The parameters $\gamma,\alpha$ chosen in Figures \ref{fig:smallk-ratio} and \ref{fig:largek-ratio} do not satisfy the gravitational wave speed constraint. Imposing this constraint would imply $\tilde{\gamma} \approx 0$. Under the gravitational wave constraint it is expected that in general the difference of the gravitational slip compared to LCDM completely disappears in the Horndeski-like model unless we allow for a $k$-dependence of the gravitational wave speed. Extensions of the Lagrangian to a broader subclass of Horndeski theory would therefore be interesting to study and would allow for studying the gravitational slip to entropy ratio as we did, which could perhaps also introduce additional bounds such as the $\eta/S \geq 1/4\pi$ found for black holes \cite{Feng:2015oea}. The entropy bounds would then also allow to constrain theoretically allowed parameter space further in a less trivial way. Observationally the gravitational slip can of course be constrained much more efficiently via galaxy clusters, however the goal in this work is to focus on theoretical bounds. \\

\begin{figure}
    \centering
    \includegraphics[width=1\linewidth]{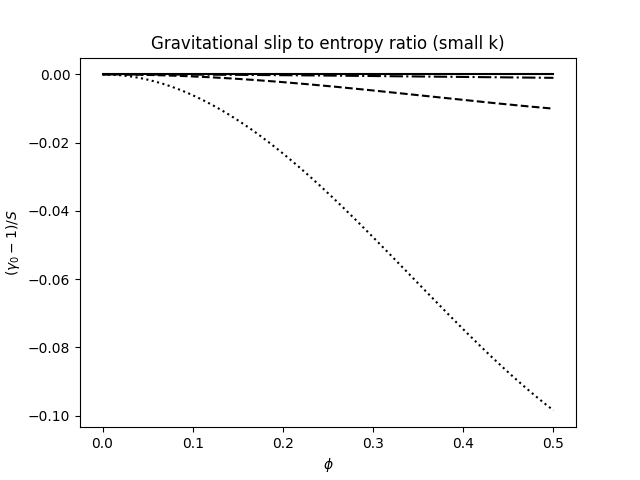}
    \caption{$(\gamma_0-1)/S$ versus $\phi$ for different values of $\gamma$ and fixed $\alpha = \Lambda = 1$. The values of $\gamma$ plotted are: $\gamma = 0.1$ (dotted), $\gamma = 0.01$ (dashed), $\gamma = 0.001$ (dashdot) and $\gamma = 0$ (solid). The plot assumes the superpotential $W(\phi) = \cos^{2/3}(3\phi/2)$,  canonical normalization of the scalar field and the equation for $\gamma_0$ in the QSA.}
    \label{fig:smallk-ratio}
\end{figure}

\begin{figure}
    \centering
    \includegraphics[width=1\linewidth]{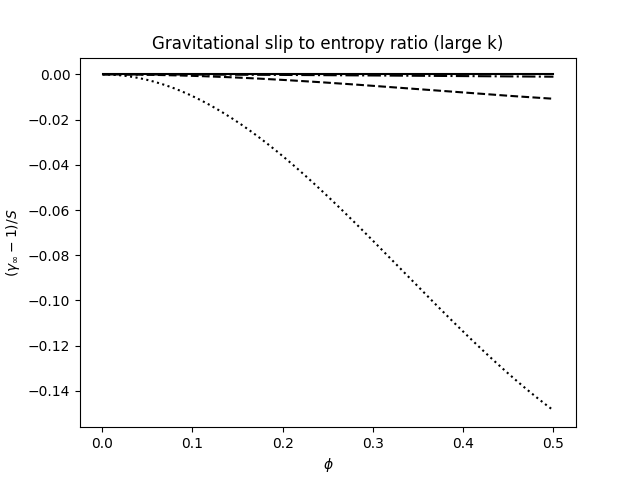}
    \caption{$(\gamma_\infty-1)/S$ versus $\phi$ for different values of $\gamma$ and fixed $\alpha = \Lambda = 1$. The values of $\gamma$ plotted are: $\gamma = 0.1$ (dotted), $\gamma = 0.01$ (dashed), $\gamma = 0.001$ (dashdot) and $\gamma = 0$ (solid). The plot assumes the superpotential $W(\phi) = \cos^{2/3}(3\phi/2)$, canonical normalization of the scalar field and the equation for $\gamma_\infty$ in the QSA.}
    \label{fig:largek-ratio}
\end{figure}

Regarding the Lagrangian in equation (\ref{eq:Horn}) in the low-energy limit and assuming the usual energy scales $\Lambda_2^4 = M_{\mathrm{pl}}^2 H_0^2$ and $\Lambda_3^3 = M_{\mathrm{pl}}H_0^2$, it is possible to compute the so-called positivity bounds in the decoupling limit $M_{\mathrm{pl}} \rightarrow \infty$ with $\Lambda_3$ fixed \cite{deBoe:2024gpf}. The result of this procedure are the bounds: $-\gamma^2 \geq 0$ and $\alpha \gamma \geq 0$. This implies that $\gamma=0$ and $\alpha \in \mathbb{R}$ arbitrary. The bound $\alpha \geq 0$ could be imposed from the point of view of the existence of a healthy Minkowski limit \cite{deBoe:2024gpf}. The results are then consistent with the finding from the entropy for particle production for the superpotential solution we considered but different from the entropy bounds coming from the apparent horizon. However, these positivity bounds are evaluated strictly speaking on the Minkowski background, but the usual assumption is to transport them to the FRW background. The transportation may not in general valid. Furthermore, this approach relies on an agnostic approach on the UV completion (except some general required properties such as unitarity and causality) of the low-energy theory and therefore it was complicated to check the validity of the transportation of the bounds between the different backgrounds. Studying the consequences of the different entropy contributions is much easier and will provide us with additional bounds allowing to constrain the allowed parameter space further facilitating in the search of finding viable Horndeski models. These entropy bounds are furthermore evaluated under the assumption of a FRW universe rather than flat spacetime, which is a motivation to study the bounds that come from entropy considerations. 

\section{Conclusions and future directions} \label{sec:conclusions}
In this paper we considered stability conditions of the scalar sector of the Horndeski-like theory using the ADM formalism, extending the work \cite{Santos:2024ynr} which considered the tensor sector. We computed the scalar stability conditions for a general cosmological background under the assumption that we consider late times so that matter and radiation can be neglected. The result of this part of the work is that there exists a choice for the superpotential $W(\phi)=\cos^{2/3}(3\phi/2)$ with $\alpha \geq 0$ and that the corresponding scalar stability constraints impose the bound $\gamma/\alpha \lessapprox 1+1.25\alpha$ compared to the previously found tensor stability constraint, which gave the bound $\gamma/\alpha \lessapprox 1.75$. \\

We studied the entropy associated with the particle production \cite{Yu:2018qzl,Prigogine:1988jax,Kaur:2021dix,Dai:2021dog,Cipriano:2023yhv}. Under the assumption that the entropy can only increase with time we could derive so-called entropy bounds from this analysis. The entropy bound for the superpotential $W(\phi)=\cos^{2/3}(3\phi/2)$ reduced to the bound $\gamma \leq 0$ in accordance with the scalar and tensor stability constraints. A similar derivation also allowed us to derive entropy bounds from the entropy of the apparent horizon. The result from the computation of the entropy of the apparent horizon is the bound $3\alpha \leq \gamma \Lambda$ for different cosmological scenarios such as $a(t) \propto t^n$ and for the case of superpotential $W(\phi)=\cos^{2/3}(3\phi/2)$. For the solution $a(t)\propto e^{H_0(t-t_0)}$ the entropy bound from the apparent horizon was found to be trivially satisfied. The entropy bounds provide different constraints compared to the stability conditions for the allowed parameter space. They are also much easier to compute than the usual positivity bounds \cite{deBoe:2024gpf}, which often rely on the computation on flat spacetime rather than a FRW universe. \\

We also investigated the gravitational slip to entropy ratio. The result from the analysis is that it can indeed give some insight in the evolution of the entropy and gravitational slip during some part of cosmic evolution. However, for the specific model we considered we found that for the gravitational slip goes to the LCDM value quickly once $\gamma \sim 0.01$. The idea however is much more general since if it would be possible in a future work to include more terms in the Horndeski-like Lagrangian and hence by such extending the generality of the action to have more freedom in the usual $G_i$ functions, then entropy bounds 
and the gravitational slip to entropy ratio could be derived in a more general setting. A procedure as in \cite{Padilla:2012ze} could allow to find the boundary term in such a context. Especially given the fact that in the action considered in this work for a general $\phi(t)$ the value of $\gamma$ is constrained severely by the gravitational wave constraint. However, the philosophy of the approach in this work is much broader, being able to include more terms in the Horndeski-like theory will be interesting because of this reason. The entropy of the spacetime itself has not been taken into account by our analysis and could be investigated as well in a future work. The correct entropy bounds should be placed on the total entropy rather than the individual contributions, however for this the entropy of the spacetime itself should be incorporated first. Next, the stability conditions can be investigated for general cosmological solutions involving the behavior during the entire cosmic evolution rather than focusing on just a phase that describes cosmic acceleration. And it would be interesting to investigate it for general superpotentials including a non-trivial additional scalar field $\chi$ and to include a non-zero perturbation $\delta \chi$. It would be interesting to find analytically cosmological solutions of equation (\ref{eq:superp}). And also the gravitational slip and entropy production could be studied in the presence of a non-trivial additional field $\chi$. Another direction would be to consider the non-linear regime of the scalar perturbations for the given action and correspondingly the spherical collapse model in such a setting describing for instance the formation of halos in the universe via the computation of quantities such as the halo mass function and non-linear matter power spectrum. \\

\section*{Acknowledgements} \label{sec:acknowledgements}
We thank Thomas Colas for a useful discussion about the idea of replacing positivity bounds by entropy bounds and Gen Ye for useful discussions about \texttt{EFTCAMB}. We thank the anonymous referee for the useful comments on the manuscript. DdB acknowledges support from the NWO and the Dutch Ministry of Education, Culture and Science (OCW) (through NWO VIDI Grant No. 2019/ENW/00678104 and ENW-XL Grant OCENW.XL21.XL21.025 DUSC) and from the D-ITP consortium.

\end{document}